\documentclass{elsart}

\usepackage{graphicx}
\usepackage{dcolumn}
\usepackage{bm}
\usepackage{amssymb}

\begin{document}

\begin{frontmatter}

\title{Quantum Nernst Effect}

\author[nakamura]{Hiroaki Nakamura}
\ead{nakamura@tcsc.nifs.ac.jp}
\author[hatano]{Naomichi Hatano}
\author[shirasaki]{Ry\=oen Shirasaki}

\address[nakamura]{Theory and Computer Simulation Center,
National Institute for Fusion Science,
Oroshi-cho 322-6, Toki, Gifu 509-5292, Japan}
\address[hatano]{Institute of Industrial Science,
University of Tokyo,
Komaba 4-6-1, Meguro, Tokyo 153-8505, Japan}
\address[shirasaki]{Department of Physics, Yokohama National University,
Tokiwadai 79-5, Hodogaya-ku, Yokohama 240-8501, Japan}

\date{\today}

\begin{abstract}
It is theoretically predicted that the Nernst coefficient is strongly suppressed and the thermal conductance is quantized in the quantum Hall regime of the two-dimensional electron gas.
The Nernst effect is the induction of a thermomagnetic electromotive force in the $y$ direction under a temperature bias in the $x$ direction and a magnetic field in the $z$ direction.
The quantum nature of the Nernst effect is analyzed with the use of a circulating edge current and is demonstrated numerically.
The present system is a physical realization of the non-equilibrium steady state.
\end{abstract}
\end{frontmatter}


\section{Introduction}
\begin{figure}[b]
\begin{center}
\includegraphics[width=0.35\textwidth,clip]{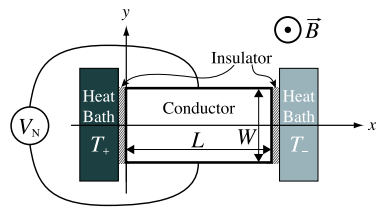}
\end{center}
\caption{A setup for observation of the Nernst effect.
The Nernst voltage $V_\mathrm{N}$ is defined as such that it is positive when the voltage of the upper edge is higher than the voltage of the lower edge.}
\label{Fig.nernst}
\end{figure}

The (adiabatic) Nernst effect in a bar of conductor is the generation of a voltage difference in the $y$ direction under a magnetic field in the $z$ direction and a temperature bias in the $x$ direction (Fig.~\ref{Fig.nernst}).
Each of the left and right ends of the conductor is attached to a heat bath with a different temperature, $T_+$ on the left and $T_-$ on the right.
An electric insulator is inserted in between the conductor and each heat bath, so that only the heat transfer takes place at both ends.
(There is no contact on the upper and lower edges.)
A constant magnetic field $B$ is applied in the $z$ direction.
Then the Nernst voltage $V_\mathrm{N}$ is generated in the $y$ direction. 
(In what follows, we always put $\Delta T\equiv T_+-T_->0$ and $B>0$.)

A classical-mechanical consideration on this thermomagnetic effect yields the following:
a heat current flows from the left end to the right end because of the temperature bias;
the electrons that carry the heat current receive the Lorentz force from the magnetic field and deviate to the upper edge;
then we have $V_\mathrm{N}<0$.
The Nernst coefficient is defined as
\begin{equation}\label{eq1}
N\equiv \frac{V_\mathrm{N}/W}{B \nabla_x T},
\end{equation}
where the temperature gradient is given by $\nabla_x T=-\Delta T/L$ with $W$ and $L$ being the width and the length of the conductor bar.
The above naive consideration gives a positive Nernst coefficient.
In reality, the Nernst coefficient can be positive or negative, depending on the scattering process of electrons.

The Nernst effect was extensively investigated in the 1960's\cite{67Harman}
because of a possible application to conversion of heat to electric energy.
The investigation on the energy conversion was eventually abandoned,
since induction of the magnetic field cost lots of energy in those days. 
The Nernst effect, however, has recently seen renewed interest\cite{94Yam2,99Nakamura,04Hasegawa};
improvement of the superconducting magnet has led to more efficient induction of a strong magnetic field.
This is a background of recent studies on the Nernst effect at temperatures higher than the room temperature.

In the present Letter, we direct our attention to the Nernst effect in the regime of the ballistic conduction, that is, the Nernst effect of the two-dimensional electron gas in semiconductor heterojunctions at low temperatures, low enough for the mean free path to be greater than the system size.
Using a simple argument on the basis of edge currents\cite{82Halperin}, we predict that, when the chemical potential is located between a pair of Landau levels,
(i) the Nernst coefficient is strongly suppressed and (ii) the thermal conductance in the $x$ direction is quantized.

Incidentally, the physical state of the present system is a realization of the non-equilibrium steady state (NESS), a new concept much discussed recently in the field of non-equilibrium statistical physics\cite{Ruelle,Ogata}.
The non-equilibrium steady state is almost the first quantum statistical state far from equilibrium that can be analyzed rigorously;
it consists of a couple of independent currents with different temperatures.
Since the study of the non-equilibrium steady state has been almost purely mathematical, we consider it valuable to give it a physical realization.

\section{Predictions}
\begin{figure}[b]
\begin{center}
\includegraphics[width=0.4\textwidth,clip]{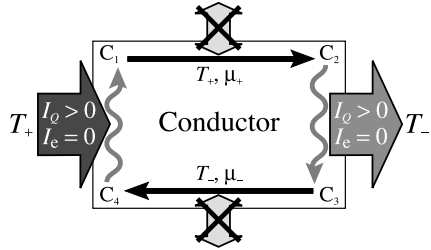}
\end{center}
\caption{A schematic view of the dynamics of electrons in a Hall bar under the setup for the Nernst effect.}
\label{fig.idea}
\end{figure}

Let us first briefly explain our basic idea (Fig.~\ref{fig.idea}).
Since there is no input or output electric current, an edge current circulates around the Hall bar when the chemical potential is in between neighboring Landau levels.
The edge current along the left end of the bar is in contact with the heat bath with the temperature $T_+$ and equilibrated to the Fermi distribution $f(T_+,\mu_+)$ with the temperature $T_+$ and a chemical potential $\mu_+$ while running from the corner C$_4$ to the corner C$_1$.
Since the upper edge is not in contact with anything, the edge current there runs ballistically, maintaining the Fermi distribution $f(T_+, \mu_+)$ all the way from the corner C$_1$ to the corner C$_2$.
It then encounters the other heat bath with the temperature $T_-$ and equilibrated to the Fermi distribution $f(T_-, \mu_-)$ while running from the corner C$_2$ to the corner C$_3$.
The edge current along the lower edge runs ballistically likewise, maintaining the Fermi distribution $f(T_-, \mu_-)$ all the way from the corner C$_3$ to the corner C$_4$.
(The circulating edge current constitutes a physical realization of the non-equilibrium steady state\cite{Ogata}.)
The Nernst voltage $V_\mathrm{N}=\Delta\mu/\mathrm{e}\equiv(\mu_+-\mu_-)/\mathrm{e}$ is thus generated, where $\mathrm{e}(<0)$ denotes the charge of the electron.

First, the difference in the chemical potential, $\Delta\mu$, is of a higher order of the temperature bias $\Delta T$, because the number of the conduction electrons is conserved.
The Nernst coefficient (\ref{eq1}), or
\begin{equation}\label{eq10}
N=\frac{1}{|\mathrm{e}|B}\frac{L}{W}\frac{\Delta\mu}{\Delta T},
\end{equation}
hence vanishes as a linear response.
Second, the heat current $I_Q$ in the $x$ direction is carried ballistically by the edge current along the upper and lower edges.
The edge current does not change much when we vary the magnetic field $B$ as long as the chemical potential stays between a pair of neighboring Landau levels.
The heat current hence has quantized steps as a function of $B$.

\section{Two-dimensional electron gas}
\begin{figure}
\begin{center}
\includegraphics[width=0.35\textwidth,clip]{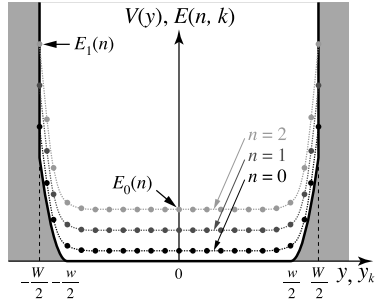}
\end{center}
\caption{A schematic view of the transverse confining potential $V(y)$, on which the structure of the Landau levels is superimposed.
The central part of the potential, $|y|<w/2$, is flat, whereas the potential edges $w/2<|y|<W/2$ may have some curvatures.}
\label{Fig.vy}
\end{figure}

We now describe the above idea explicitly.
In order to fix the notations, we begin with the basics of the two-dimensional electron gas in a magnetic field.
The dynamics of the two-dimensional electron gas is described by the Schr\"{o}dinger equation
with the vector potential $(-By,0)$ and the confining potential $V(y)$
shown schematically in Fig.~\ref{Fig.vy}.
We can express the eigenfunction 
in the form of variable separation:
$\Psi(x,y) = \mathrm{e}^{\mathrm{i}kx} \chi_k(y)/\sqrt{L}$,
where
$k = 2\pi j/L$
with an integer $j$.
The transverse part $\chi_k(y)$ is an eigenfunction of the equation
${\mathcal H}_k\chi_k(y)=E\chi_k(y)$,
where
\begin{equation}\label{ham.eq}
{\mathcal H}_k\equiv\frac{p_{y}^2}{2 m }+ \frac{m \omega_\mathrm{c}^2 }{2} \left( y - y_k \right)^2 +V(y)
\end{equation}
with
$\omega_\mathrm{c}\equiv|\mathrm{e}|B/m$
and
$y_k\equiv \hbar k/(|\mathrm{e}|B)$. 
We label the discrete eigenfunctions with an integer $n$.
The whole solution is then given by
$\Psi_{n,k}(x,y)=\mathrm{e}^{\mathrm{i}kx}\chi_{n,k}(y)/\sqrt{L}$
with
$E=E(n,k)$.
As is schematically shown in Fig.~\ref{Fig.vy}, the eigenvalue $E(n,k)$ in fact scarcely depends on $k$ in the bulk, where the confining potential $V(y)$ is flat\cite{82Halperin,84MacDonald}.

The Hamiltonian~(\ref{ham.eq}) shows that an eigenfunction with the $x$ component of the momentum, $\hbar k$, is centered around $y=y_k\propto k$.
In other words, the state in the upper half of the Hall bar has a current in the positive $x$ direction, while the one in the lower half has a current in the negative $x$ direction.
The velocity of the electron in the state $\Psi_{n,k}$ is
\begin{equation}\label{vdef.eq}
v\left( n,k \right) 
 = \frac{1}{\hbar} \frac{\partial E(n,k)}{\partial k },
\end{equation} 
which remains finite only near the upper and lower edges.
These are the edge currents.

\section{Electric and heat currents}
Now we write down the electric current $I_\mathrm{e}$ and the heat current $I_Q$ in the $x$ direction carried by electrons.
(Note that we will put $I_\mathrm{e}=0$ in the bottom line, observing the boundary conditions in Fig.~\ref{fig.idea}.)
The currents are given by
\begin{equation}\label{I.eq}
I_\mathrm{e}=\langle \mathrm{e}v\rangle
\quad\mbox{and}\quad
I_Q=\langle(E-\mu)v\rangle
\end{equation}
with the thermal average
\begin{equation}
\langle\cdots\rangle=\frac{1}{\pi} \sum_{n=0}^{\infty}
  \int_{-k_\mathrm{m}}^{k_\mathrm{m}}     
 \cdots
 f_{n,k}\left(T(y_k),\mu(y_k) \right)\mathrm{d}k,
\end{equation}
where we made the summation over $k$ to the momentum integration.
The integration limits $\pm k_\mathrm{m}$ are the maximum and minimum possible momenta.
The function $f_{n,k}$ denotes the Fermi distribution $f(T,\mu)=\{1+\exp[(E-\mu)/(k_\mathrm{B}T)]\}^{-1}$ with $E=E(n,k)$.
The layout in Fig.~\ref{fig.idea} yields $(T(y_k),\mu(y_k))=(T_+,\mu_+)$ for 
the upper edge states and $(T_-,\mu_-)$ for 
the lower edge states.

We transform $I_\mathrm{e}$
with the use of eq.~(\ref{vdef.eq}) as
\begin{equation}\label{I2.eq}
I_\mathrm{e}
 = \frac{\mathrm{e}}{\pi\hbar} \sum_{n=0}^{\infty} 
      \int_{E_0(n)}^{E_1(n)} 
	  \left[ f\left(T_+,\mu_+\right)-f\left(T_-,\mu_-\right) \right]
	  \mathrm{d}E.
\end{equation}
The integration limits are now $E_0(n)=E(n,0)$ and $E_1(n)=E(n,k_\mathrm{m})$ (see Fig.~\ref{Fig.vy}).
In order to compute the linear response, we here put
$T_\pm=T\pm\Delta T/2$ and $\mu_\pm=\mu\pm\Delta \mu/2$
with $\Delta T \ll T$ and $\Delta \mu \ll \mu$.
By expanding eq.~(\ref{I2.eq}) with respect to $\Delta T$ and $\Delta\mu$, we have
\begin{equation} \label{I3.eq}
I_\mathrm{e} 
  \simeq \frac{\mathrm{e}}{\pi\hbar} \left[ \Delta\mu \sum_{n=0}^{\infty} A_0(n)
      +k_\mathrm{B}\Delta T \sum_{n=0}^{\infty} A_1(n)  \right],
\end{equation}
and similarly
\begin{equation}\label{Q3.eq}
I_Q \simeq \frac{k_\mathrm{B}T}{\pi\hbar} \left[  \Delta \mu \sum_{n=0}^{\infty} A_1(n)
      +k_\mathrm{B}\Delta T \sum_{n=0}^{\infty} A_2(n)  \right],
\end{equation}
where
\begin{equation}\label{An.eq}
A_\nu(n)\equiv\int_{x_0(n)}^{x_1(n)}
       \frac{x^\nu\mathrm{d}x}{4 \cosh^2 \left( x/2 \right)}
\end{equation}
with
$x_i (n) \equiv \left(E_i(n)-\mu\right)/(k_\mathrm{B}T)$. 
The integral~(\ref{An.eq}) can be carried out explicitly for $\nu=0,1,2$.

Since there is no input or output current in the setup in Fig.~\ref{Fig.nernst}, we put $I_\mathrm{e}\equiv0$ in eq.~(\ref{I3.eq}), relating $\Delta\mu$ with $\Delta T$.
as
\begin{equation}\label{muT.eq}
\Delta \mu = -\frac{\sum_n A_1(n)}{\sum_n A_0(n)}k_\mathrm{B}\Delta T.
\end{equation}
We thereby arrive at the Nernst coefficient~(\ref{eq10}) as
\begin{equation}\label{Nernstresult}
N  = -\frac{k_\mathrm{B}}{|\mathrm{e}|B} \frac{L}{W} \frac{\sum_n
                  A_1(n)}{\sum_n A_0(n)}.
\end{equation}
The heat current~(\ref{Q3.eq}) 
yields the thermal conductance as
\begin{equation}\label{Conductanceresult}
G_Q\equiv \frac{I_Q}{\Delta T}
=\frac{{k_\mathrm{B}}^2T}{\pi\hbar} \left[  \sum_n A_2(n)- \frac{\left(\sum_n A_1(n)\right)^2 }{ \sum_n A_0(n) }
                         \right]. 
\end{equation}

\textit{Low-temperature limit}:
In the low-temperature limit, the upper and lower limits of the integration in eq.~(\ref{An.eq}) goes to $\pm\infty$, depending on their signs.
First, in the usual experimental situation, the confining potential at its edges (of the order of eV) is considerably higher than the chemical potential (of the order of meV);
hence we assume $E_1(n)>\mu$ for all $n$.
The upper integration limit $x_1(n)$ thus always goes to $+\infty$ as $T\to 0$.
Next, suppose that the chemical potential is located in between the bottom of the $(M-1)$th Landau level and the bottom of the $M$th one.
The lower integration limit $x_0(n)$ goes to $+\infty$ for $n\geq M$ and the integral vanishes as $T\to0$.
The integral can survive only for $n\leq M-1$, for which the lower integration limit $x_0(n)$ goes to $-\infty$ as $T\to0$, yielding
$A_0(n)=1$,
$A_1(n)=0$,
and $A_2(n)=\pi^2/3$.
Thus we arrive at the predictions
\begin{equation}\label{eq.lowT}
N=0\qquad\mbox{and}\qquad
\frac{G_Q}{T}=\frac{\pi{k_\mathrm{B}}^2}{3\hbar}M
\end{equation}
when the chemical potential is located in between the bottoms of a pair of the neighboring Landau levels.

\section{Numerical demonstration}
Let us demonstrate the above by adopting the following confining potential\cite{92Komiyama}:
\begin{equation}
 V(y) = 
  \left\{
    \begin{array}{ll}
      0           & \mbox{for}\quad |y| \leq \frac{w}{2} , \\
      \frac{m \omega_0^2}{2} \left( |y| - \frac{w}{2} \right)^2
             & \mbox{for}\quad \frac{w}{2}< |y| <\frac{W}{2}.
    \end{array}
  \right.
  \label{v.eq}
\end{equation}
The eigenvalues are well approximated in each region of eq.~(\ref{v.eq}) by tentatively regarding that the potential there continues for all $y$\cite{82Halperin}.
This approximation is valid because, in the parameter range that we use below,
each eigenfunction is well localized in the $y$ direction and insensitive to the potential elsewhere.
The mismatch of the approximated eigenvalue
at $|y|=w/2$ is much smaller than the eigenvalue itself
in the parameter range below.\cite{92Komiyama}
Furthermore, the eigenvalue is shifted
right on the edges because of the potential walls\cite{84MacDonald}.
This shift contributes only to a shift of $x_1(n)$, which is irrelevant as $x_1(n)$ is virtually infinite anyway.

We set the parameters as follows:
the effective mass is $m=0.067 m_0$ for GaAs, where $m_0$ is the bare mass of the electron;
the sample size is $L=20\mu$m and $W=20\mu$m (less than the mean free path at low temperatures\cite{Tarucha}) with $w=16\mu$m;
the confining potential is given by $V(\pm W/2) = 5.0$eV, the work function of GaAs;
the chemical potential is $\mu=15$meV, which means the carrier density $n_\mathrm{s}=4.24\times 10^{15}\mathrm{m}^{-2}$.

\begin{figure}[t]
\begin{center}
\includegraphics[width=0.45\textwidth]{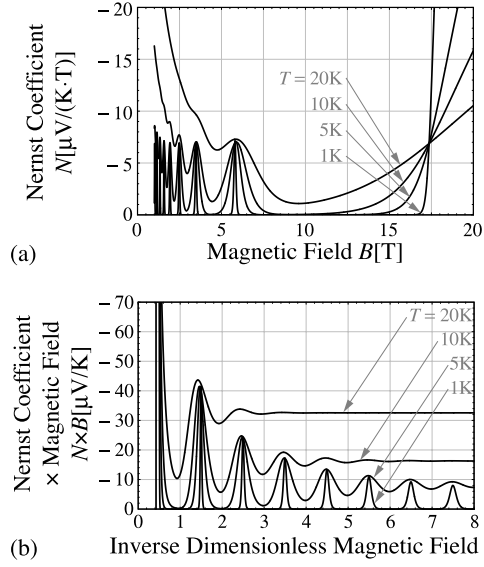}
\end{center}
\caption{The magnetic-field dependence of the adiabatic Nernst coefficient at $T=1,5,10$ and $20$K for $1\mathrm{T}\leq B\leq 20\mathrm{T}$;
(a) the raw data and (b) a scaling plot of $N \times B$ against $m\mu/\hbar|\mathrm{e}|B$.}
\label{Fig.coeff}
\end{figure}

\begin{figure}[t]
\begin{center}
\includegraphics[width=0.45\textwidth]{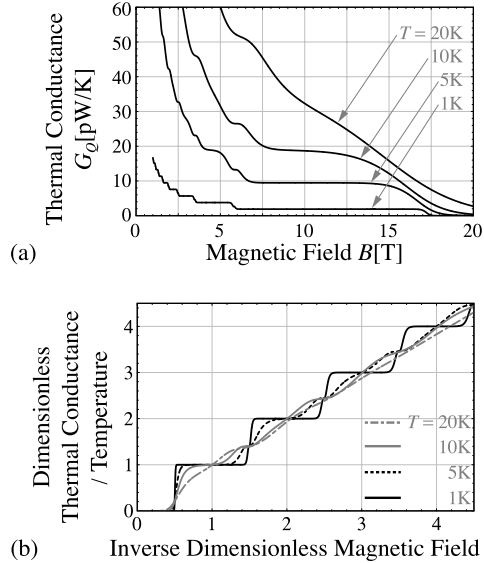}
\end{center}
\caption{The magnetic-field dependence of the thermal conductance at $T=1,5,10$ and $20$K for $1\mathrm{T}\leq B\leq 20\mathrm{T}$;
(a) the raw data and (b) a scaling plot of $G_Q/T\times3\hbar/\pi{k_\mathrm{B}}^2$ against $m\mu/\hbar|\mathrm{e}|B$.}
\label{q.fig}
\end{figure}

Using these values, we obtained the adiabatic Nernst coefficient~(\ref{Nernstresult}) as in Fig.~\ref{Fig.coeff}
and the thermal conductance~(\ref{Conductanceresult}) as in Fig.~\ref{q.fig}.
We see that our predictions~(\ref{eq.lowT}) are indeed realized at low temperatures.
We also note that the Nernst coefficient is negative in the present case.

\section{Thermopower}
We comment here on the study of \lq\lq the thermopower of the two-dimensional electron gas" in the 1980's.
Although the mathematics resembles ours, the physical situation is much different.

The theoretical study of the thermopower\cite{82Zelenin,82Girvin,83Streda,84Jonson} considered edge currents with different temperatures on opposing edges with adiabatic boundary conditions.
This situation is indeed the same as the upper and lower edge currents in Fig.~\ref{fig.idea}.
Therefore, Eq.~(\ref{muT.eq}) is the same as the relation that they derived between the temperature difference and the chemical-potential difference of the two edge currents as a longitudinal response.
The theory, however, did not come up with any experimental setup that could realize the above situation.
The experiments\cite{84Obloch,86Obloch,86Davidson,86Fletcher}, in fact, were carried out in a situation where edge currents were in contact with heat baths with different temperatures;
that is, the experimental study considered non-adiabatic boundary conditions on the upper and lower edges in Fig.~\ref{fig.idea}.
In spite of this essential difference did  they compare the theoretical predictions with the experimental results!

In comparison with the study in the 1980's, our novel point is to propose a way of realizing the situation where edge currents with different temperatures coexist, by considering the circulating edge current in Fig.~\ref{fig.idea}.
Thanks to the setup of the Nernst effect,  the \textit{external} temperature gradient in the $x$ direction appears in the \textit{internal} temperature gradient in the $y$ direction.
As yet another point, the experiments in the 1980's were 
out of the ballistic regime.
%
The experimental technique has been developed remarkably since the 1980's, when it was almost impossible to achieve the ballistic transport.
In fact, Hasegawa and Machida\cite{nernst.experiment} are planning an experiment in the situation of the present theory.
The present predictions have become worthwhile in the light of new technology.

\section{Summary}
We predicted a prominent quantum effect of the two-dimensional electron gas, which is closely analogous to the quantum Hall effect.
As long as the chemical potential stays in between the bottoms of the neighboring Landau levels, the quantized nature of the edge currents suppresses the Nernst coefficient and fixes the thermal conductance.
We also noted that the system is a physical realization of the non-equilibrium steady state.

The precise forms of the peaks in Fig.~\ref{Fig.coeff} and the risers of the steps in Fig.~\ref{q.fig} may be different from the reality.
This is because our argument using the edge currents is not applicable when the chemical potential coincides with the bottom of a Landau level, namely when $\mu=(n+\frac{1}{2})\hbar\omega_\mathrm{c}$, or $1/B=(n+\frac{1}{2})\hbar|\mathrm{e}|/m\mu$.
There the heat current is carried by bulk states as well as the edge states.
We then have to take account of impurities and possibly electron interactions\cite{future}.

Comments on other approaches to the quantum Nernst effect are in order.
Kontani derived\cite{02Kontani,03Kontani} by the Fermi liquid theory general expressions of the Nernst coefficient and the thermal conductivity of strongly correlated electron systems such as high-$T_\mathrm{c}$ materials.

\section*{Acknowledgments}
The authors express sincere gratitude to Dr.~Y.~Hasegawa and Dr.~T.~Machida for useful comments on experiments of the Nernst effect and the quantum Hall effect.
This research was partially supported by the Ministry of Education, Culture, Sports, 
Science and Technology, 
Grant-in-Aid for Exploratory Research, 2005, No.17654073.
N.H.~gratefully acknowledges the financial support by Casio Science Promotion Foundation and the Sumitomo Foundation.

\end{document}